\documentstyle[11pt,newpasp,twoside,psfig]{article}
\begin{document}

\title{A physicist's view of stellar dynamics: dynamical instability of stellar systems
}
\author{V.G.Gurzadyan}
\affil{Yerevan Physics Institute, Armenia and
Department of Physics, University of Rome 'La Sapienza', Italy}
 
\setcounter{page}{111}
\index{Gurzadyan, V.G.}

\begin{abstract}
I argue that the widely adopted framework of stellar dynamics
survived since 1940s, is not fitting the current knowledge on non-linear systems. 
Borrowed from plasma physics when several fundamental features of perturbed non-linear systems were unknown,
that framework ignores the difference in the role of perturbations in two
different classes of systems, in plasma with Debye screening and gravitating systems with no screening.
Now, when the revolutionary role of chaotic effects is revealed even
in planetary dynamics i.e. for nearly integrable systems,  one would expect that for stellar systems, i.e. non-integrable systems, their role have to be far more crucial. 
Indeed, ergodic theory tools already enabled to prove that spherical stellar systems are exponentially instable
due to N-body interactions, while the two-body encounters, contrary to existing belief, are not the 
dominating mechanism of their relaxation. 
Chaotic effects distinguish morphological and other properties of galaxies. 
Using the Ricci curvature criterion, one can also show 
that a central massive object (nucleus) makes the N-body gravitating system more instable (chaotic), 
while systems with double nuclei are even more instable than those with a single one.
\end{abstract}

\section{On the current framework of stellar dynamics}

Since this is a Joint Discussion at IAU General Assembly, I allow myself to start from some general but also provocative remarks; for detailed refs see (Allahverdyan, Gurzadyan 2002).
The current framework of stellar dynamics is the one summarized in Chandrasekhar's book of 1942. That framework was borrowed earlier from the plasma physics when many features of perturbed non-linear systems were unknown. This resulted in the ignorance of the drastic difference in the role of perturbations for two different classes of systems, plasma and gravitating systems: with Debye screening and justified cutoff of perturbations for the former, and long range interaction and no screening for the latter. Correspondingly, the two-body (Rutherford) scatterings, i.e. neglecting the perturbations of other particles of the system, were {\it a priori} assumed as the universal mechanism of relaxation of stellar systems,\footnote{The two-body relaxation
is postulated also in kinetic (diffusion coefficients) and other approaches to stellar dynamics.} even though it failed to explain even elliptical galaxies, the most well-mixed systems in the Universe, predicting time scales exceeding their age
(Zwicky paradox).     

Does the framework of stellar dynamics fit the current knowledge of the non-linear systems?
To address this question maximally briefly, I will concentrate only on nearly integrable systems, linked with planetary dynamics and on non-integrable ones, i.e. on the class of systems, the stellar systems belong to.

{\it Nearly integrable systems}. I will illustrate the scale of changes occured since 1940s mentioning two works, the Kolmogorov theorem (1954) and Fermi-Pasta-Ulama (FPU, 1955) experiment.
Done practically at the same time, at the different sides of the iron curtain, in Moscow and Los Alamos, both works came to contradict the views held almost during half a century, since Poincare's theorem on the perturbed Hamiltonian systems. Kolmogorov theorem (now the main theorem of Kolmogorov-Arnold-Moser (KAM) theory) had tremendous impact on the study of dynamical systems, including the dynamics of the Solar system. FPU has inspired numerous studies (including the discovery of solitons), however in spite of much efforts, the dynamics of that 64-particle nonlinearly interacting one-dimensional system remains not completely understood up to now. Maybe this lesson has to be taken into account also for stellar dynamics.      

{\it Non-integrable systems}.
After the discovery of the metric invariant by Kolmogorov (1958), KS-entropy, and introduction of K-systems (Kolmogorov, 1959), 'an unexpected discovery' (to quote Arnold) was made in 1960s
(Anosov, Sinai, Smale) on the structural stability of exponentially instable systems. The emerged ergodic theory provided the classification of non-integrable systems by their statistical properties, with corresponding criteria and tools, though the latter not always were easy to apply for a given physical system. Those achievements enabled to attack several long standing problems such as the relaxation of Boltzmann gas, and served as the framework for the study of chaos during the following decades. 

KAM theory ideas when applied in planetary dynamics by Laskar, Tremaine and others 
revealed the fundamental role of chaos in the evolution of the Solar system, predicting the possible escape of Mercury from its orbit due to chaotic variations of the eccentricity, chaotic variations of the obliquity of Mars and the stabilization of the same effect by the Moon in the case of Earth (Laskar). So, if already for planetary systems i.e. for nearly integrable systems, the chaotic effects due to small perturbations of planets lead to such unexpected results, how can stellar systems, i.e. non-integrable many-dimensional systems avoid the influence of chaos due to the perturbations of N particles?   

Ergodic theory tools were applied in stellar dynamics in (Gurzadyan, Savvidy 1984, below GS), where the spherical systems were shown to be exponentially instable systems and the time scale of tending to microcanonical state (the relaxation time) was estimated using the standard Maupertuis reparameterization for the geodesic flow, as follows from the theorems of ergodic theory.\footnote{The Maupertuis reparameterization of the affine parameter (time) of the geodesics corresponds to the conservation of total energy of the system. Numerical experiments without such reparameterization performed first by Miller in 1964, repeated later by Heggie, Hut, Kandrup and others, therefore violate the energy conservation condition and have no link with the mentioned statistical properties and relaxation of the system.} 
More important, the results in GS and in (Pfenniger 1986) (using the Lyapounov formalism) came to reveal that, the
plasma analogy in the linear (!) sum of scattering angles at subsequent two-body scatterings is irrelevant for a long-range non-linear system's dynamics, and N-body scatterings do contribute to the statistical properties and hence in the relaxation of stellar systems.  
Particularly, the formula derived in GS for the relaxation time scale due to non-linear effects provided enough time for the relaxation of elliptical galaxies.
 By now that formula is supported by numerical simulations, alternative theoretical derivation, observational data on globular clusters and elliptical galaxies; see refs in (Allahverdyan, Gurzadyan 2002). 

There are preliminary indications from deep surveys on the existence of elliptical galaxies at redshifts $z > 4$, i.e. of 10 per cent of their present age. If confirmed, this fact would moreover require more rapid mechanism of relaxation than the two-body one.

The chaotic effects are not only responsible for the relaxation and evolution
of globular clusters and elliptical galaxies, but also they are indicators of the morphological type and other properties of galaxies.

How many decades are needed to realize the necessity of replacement of the 'plasma' framework of stellar dynamics and abandoning of the two-body relaxation myth?    

\section{Relative instability of stellar systems}

 I will now briefly discuss the problem of relative instability of stellar systems,  concentrating particularly on the role of central massive objects, in view of recent progress in their studies in the cores of galaxies and globular clusters. The results are obtained by means of the above mentioned ergodic theory formalism.

In accord to the criterion introduced in (Gurzadyan, Kocharyan 1988) among two systems the one with smaller negative Ricci curvature $r_u$  
has to be considered as more instable. For N-body gravitating systems the Ricci curvature equals
\begin{equation}
r_u(s)=- \frac{(3N-2)}{2} \frac{W_{ik}u^iu^k}{W}+ \frac{3}{4}(3N-2)
\frac{(W_iu^i)^2}{W^2}-
\frac{(3N-4)}{4} \frac{|\nabla W|^2}{W^3},
\end{equation}
where
$$
W = E - V, \quad V = - G\sum_{i<j}^{N} m_i m_j/r_{ij},
$$
and
$$
W_i = \frac{\partial W}{\partial r^i}, \quad
W_{ik}=\frac{\partial ^2W}{\partial r^i\partial r^k}, \quad
|\nabla W|^2= \sum_{i}(\frac{\partial W}{\partial r_i})^2,
$$
$m_i$ denote the masses, $u$ is the velocity of geodesics with affine parameter $s$ in the configuration space. The minimal values of the Ricci curvature have to be compared within given interval of the affine parameter. The contribution of direct impacts of stars, i.e. when two stars get the same coordinates, is neglected in Eq.(1), as they are rare events for real stellar systems.

The idea is based on the average description of the exponential deviation of geodesics in the configuration space with sign-indefinite curvature tensor. The latter condition appears to be fulfilled for N-body gravitating systems thus indicating the diversity of possible configurations with very different properties, from semi-regular planetary systems to mixing spherical systems.
The $r_u(s)$ is related with the Ricci tensor $Ric$ by the following expression
$$
r_u(s)= \frac{{\it Ric}(u,u)}{u^2}.
$$     
This criterion has a principal difference from that of Lyapounov exponents, since provides local in time characteristics of the system and hence does not require 'infinite' iterated computations.    

Numerical experiments using this criterion have been performed for various N-body configurations by Bekarian, El-Zant, Melkonian, Kocharyan and others. It is natural to see that, for example, disk configurations with rotational momentum are more regular than spherical ones. More rigorous consideration based on Arnold's theorem (1966) on one-parametric groups of manifolds with right-invariant Riemannian metric enables one to conclude that the Galactic disk does not possess the property of mixing (and hence the corresponding relaxation time scale), as spherical stellar systems do.
 
The following classification for the systems of our interest by the increase of statistical properties is emerging from numerical experiments:

1. Spherical systems;

2. Systems with a massive central object (nucleus);

3. Systems with double nuclei.

The role of a massive center, with similar conclusions, has been studied using other methods by van Albada, Norman, Rauch, Tremaine and others. We now see that double (or binary) massive objects, like those apparently observed in galaxies Markarian 273, Arp 220, have to make the system even more chaotic, i.e. with further increase in the rate of evolution driving effects (Bekarian, Melkonian 2000).  

The topics mentioned above on the role of non-linear effects in stellar dynamics gain more importance in view of ever increasing possibilities of numerical experiments (Makino 2003).

\end{document}